\documentclass{article}

\usepackage{PRIMEarxiv}

\usepackage{natbib}
\usepackage{enumitem}
\usepackage[utf8]{inputenc} 
\usepackage[T1]{fontenc}    
\usepackage{hyperref}       
\usepackage{url}            
\usepackage{amsfonts}       
\usepackage{nicefrac}       
\usepackage{microtype}      
\usepackage{fancyhdr}       
\usepackage{graphicx}       
\graphicspath{{figs/}}     
\usepackage{subcaption}
\usepackage{float}
\usepackage{comment}

\usepackage{algorithm2e}

\usepackage{xcolor}

\usepackage{amsmath}

\usepackage{setspace}
\renewcommand{\theenumi}{\Alph{enumi}}

\title{Improving Recommendation System Serendipity Through Lexicase Selection}

\author{
  Ryan Boldi\thanks{RB and AL contributed equally} \\
  University of Massachusetts Amherst \\
  Amherst, MA\\
  \texttt{rbahlousbold@umass.edu} \\
   \And
  Aadam Lokhandwala\footnotemark[1] \\
   University of Massachusetts Amherst \\
  Amherst, MA\\
  \texttt{alokhandwala@umass.edu} \\
   \And
   Edward Annatone \\
    University of Massachusetts Amherst \\
  Amherst, MA\\
   \texttt{eannatone@umass.edu} \\
   \And
   Yuval Schechter \\
    University of Massachusetts Amherst \\
  Amherst, MA\\
   \texttt{yshechter@umass.edu} \\
   \And
   Alexander Lavrenenko \\
    University of Massachusetts Amherst \\
  Amherst, MA\\
   \texttt{alavrenenko@umass.edu} \\
   \And
   Cooper Sigrist \\
    University of Massachusetts Amherst \\
  Amherst, MA\\
   \texttt{csigrist@umass.edu} \\
}

\begin{document}
\maketitle
\begin{abstract}
Recommender systems influence almost every aspect of our digital lives. Unfortunately, in striving to give us what we want, they end up restricting our open-mindedness. Current recommender systems promote echo chambers, where people only see the information they want to see, and homophily, where users of similar background see similar content.
We propose a new serendipity metric to measure the presence of echo chambers and homophily in recommendation systems using cluster analysis. We then attempt to improve the diversity-preservation qualities of well known recommendation techniques by adopting a parent selection algorithm from the evolutionary computation literature known as lexicase selection. 
Our results show that lexicase selection, or a mixture of lexicase selection and ranking, outperforms its purely ranked counterparts in terms of personalization, coverage and our specifically designed serendipity benchmark, while only slightly under-performing in terms of accuracy (hit rate). We verify these results across a variety of recommendation list sizes. In this work we show that lexicase selection is able to maintain multiple diverse clusters of item recommendations that are each relevant for the specific user, while still maintaining a high hit-rate accuracy, a trade off that is not achieved by other methods.
\end{abstract}

\section{Introduction}

Recommender systems have become an essential aspect our daily lives. They aid us with our shopping, entertainment, research, and much more. These systems are at the heart of how we interact with technology, as well as how we interact with other people using technology. They leverage user data and specific learning algorithms to generate personalized recommendations and improve user engagement. However, recommender systems are not perfect. While they optimize to match our preferences, they end up being restricting. While they optimize to be consistent, they end up being monotonous. Recommender systems may inadvertently restrict user diversity and prevent them from exploring their preferences fully. This choice of what to optimize results in disregarding content that is exciting and interesting for users. Instead, they are shown content that is the same as what they've seen previously, or the same as what people similar to them like.

The extent to which we are shaped by the items and people we interact with has been significantly studied within the psychological and sociological literature \cite{zuckerman_2008, ECHDAA, knijnenburg_recommender_2016}. Homophily, the tendency for individuals to associate with others who share similar attributes, is one such shaping. Collaborative Filtering (CF) is built on principles very similar to homophily, relying on user-item interaction data to generate recommendations based on similarity metrics. Systems like these are responsible for shaping the content we view and choose to consume on a large scale. For this reason, it is imperative to investigate the role of recommender systems in shaping user behavior and promoting algorithmic bias.

In this work, we present a benchmark supported by the psychological and sociological literature that measures a form of serendipity in the recommendations provided to users based, on the standard metric of serendipity that is in use currently \citep{toms_serendipitous_2000}. These serendipity benchmarks determine whether items that are both useful and unexpected for users. Different to the usual recommender system serendipity measurements, our framework is directly set up to detect echo chambers and homophilous circles by analyzing a clustering of the items recommended to each user. We then present the application of selection technique from evolutionary computation known as lexicase selection that has been shown to preserve diversity in Genetic Programming runs \citep{helmuth_lexicase_2019}, which results in improved overall performance \citep{helmuth_applying_2022}. We replace the final ranking phase of a Neural Matrix Factorization (NeuMF) \cite{he_neural_2017} recommender system with a phase that uses lexicase selection to shortlist the items for each user. We find that this change results in a significant improvement in our serendipity benchmark for a variety of recommender list sizes. We find that our new system, optimized on our metric, improved the Coverage, Personalization, and Hit Rate of the vanilla recommender systems.

\section{Background and Related Work}

Homophily, the phenomenon in which individuals with similar backgrounds tend to share similar preferences, poses a significant challenge to the effectiveness of recommendation systems \citep{zuckerman_2008}. This is especially true in online settings, where it can lead to the formation of echo chambers, in which users are exposed only to information and viewpoints that reinforce their existing beliefs and preferences. To combat this phenomenon, researchers have focused on promoting serendipity, or the discovery of unexpected and diverse items \citep{toms_serendipitous_2000}. These efforts have led to the development of a range of techniques, with varying degrees of success. 

Some of the most common methods for benchmarking recommendation systems are precision and recall. Precision measures the proportion of relevant items retrieved out of all the items retrieved, while recall measures the proportion of relevant items retrieved out of all items that are relevant to the user \cite{herlocker_evaluating_2004}. Hit rate, a metric that measures the number of items correctly predicted to have been rated by users in a held-out data set, also serves as an accuracy benchmark \cite{Shani2011}. Other benchmarks, such as coverage, which measures the proportion of items for which predictions can be generated, and personalization, which assesses whether the system recommends different users similar items, also play important roles in evaluating recommendation systems \cite{Ge2010}.

Metrics measuring diversity and novelty are very relevant to this work. Diversity and novelty are important metrics for evaluating recommendation systems, but can be counterproductive if users are shown items that are not interesting to them. While these metrics often use dissimilarity measures to previous items viewed (novelty) or between items within a recommendation list (diversity), they do not consider user identity and embedding, which can lead to homophily and bias in recommendations \citep{castells_novelty_2011, adomavicius_2012_aggreagate}. Accuracy metrics in collaborative filtering and other recommendation approaches typically use similarity metrics such as similar users' preferences for similar items or the similarity between an item and previously liked items for the same user. Researchers have attempted to balance accuracy and diversity by examining the intra-list similarity (ILS) of recommended items, but often find that diveristy comes at the expense of accuracy \citep{ziegler_improving_2005, mcnee_2006_accuracyNotEnough, zhang_avoiding_2008}.

A different class of benchmarks that are intimately related to this work is measuring serendipity \cite{toms_serendipitous_2000, kaminskas_diversity_2017, herlocker_evaluating_2004}. These serendipity benchmarks find items that are both useful and unexpected. They often take into account user preferences and item features to determine the surprise a series of recommendations brings \citep{iaquinta_introducing_2008, oku_fusion-based_2011, onuma_tangent_2009, Adamopoulos_2015_unexpected}. In this work, we propose a modified serendipity metric that is sensitive to the vast ranges of interests that users have, and measures how much the ranking list spans these interests. This helps address echo chambers and homophily directly.

We propose a recommendation list generation method that prioritizes serendipity using evolutionary computation techniques. Evolutionary approaches have been used in recommender systems previously, with some works focusing on accuracy and diversity simultaneously \citep{hou_two-phase_2020, hinojosa-cardenas_multi-objective_2020, HorvathEvoReview2017}. Evolutionary computation uses individuals as candidate solutions, with diversity being a crucial component as we need a variety of individuals to select from to evolve better solutions. Quality Diversity (QD) approaches define behavioral descriptors (BDs) for diversity, and attempt to prioritize a spread over these BDs, while also increasing the quality of the solutions. Whilst effective, this usually requires human insight to define the BDs (like MAP-Elites), or ignore quality metrics entirely (like Novelty Search) \citep{cully2015robots, pugh_2015_confronting, lehman_abandoning_2011}. Instead, we propose the use of lexicase selection, a parent selection technique that has been found to spontaneously promote diversity without a reliance on explicit diversity clues \cite{spector_assessment_2012, helmuth_lexicase_2019, helmuth_effects_2016, helmuth_applying_2022}.

In this paper, we initiate our discussion by addressing the detection of echo chambers and homophily. We then propose a promising method for mitigating these issues based on lexicase selection. Lastly, we benchmark the effects of incorporating lexicase selection on diversity and accuracy.

\section{Detecting Echo Chambers and Homophily}
\label{sec:headings}

This paper proposes a new benchmark for measuring the ability of recommender systems to promote serendipity and diversity in users' recommendations. Specifically, we aim to detect the presence of echo chambers and homophily in the recommendation lists that users are shown. The focus of this benchmark is on promoting diversity that encourages users to explore outside of their comfort zone, rather than simply showing them `novel' items as done in prior work \citep{herlocker_evaluating_2004, Awati2022EnhancingDI, adomavicius_2012_aggreagate}. To achieve this, the proposed method analyzes item embeddings and user specific recommendation lists to find the number of distinct yet relevant clusters for each user. This approach measures the level of diversity that is still within the user's interest and determines whether recommendations are complementary or overlapping. This metric can be applied to static datasets and serves as a supplement to previous work on serendipity \citep{kaminskas_diversity_2017}.

\subsection{Our Proposed Serendipity Metric}
The basic motivation of this metric is the fact that a diverse list of recommendations isn't particularly useful if the individual for whom they are recommended is completely disinterested in every item. Therefore, to start, we make clusters out of item embeddings and only consider the clusters deemed relevant to the user. Relevance is determined using user history as follows: If a user has rated an item high (above a certain threshold) in some cluster, then that cluster is relevant. For example, if a user has rated a specific horror movie highly, the cluster that that movie finds itself in is deemed relevant to the user. We perform k-means clustering to create $\sqrt{|I|}$ clusters where $ |I| $ is the number of items available to be ranked.

Furthermore, any cluster that is nearby a relevant cluster is also deemed relevant. We can determine how ``nearby" two clusters are by performing a clustering on our previously determined cluster centroids, again using k-means. We will call this collection of relevant clusters $C$. We will define $n$ to be the number of relevant \textbf{unique} clusters represented by the items in $R$ where $R$ is the list of recommendations generated by some recommender system for a particular user. For example, if all the items that are recommended to a user come from the same cluster, $n$ will be 1.  We will calculate our metric as follows: 

$$\text{Serendipity Score} = \frac{n}{\min\left(|C|, |R|\right)}$$

\subsection{Intuitive Understanding}
This metric effectively calculates the maximum attainable diversity given the constraints of the movies in the recommendation list $R$ and the size of the recommendation list $k$. The metric will be high when the majority of the relevant clusters are represented in the recommendation list. This is intuitive as for serendipity we want to ensure as many distinct yet relevant clusters of items are represented as possible. It's also worth noting that we use $\min(|C|,|R|)$ in place of $|C|$ as there could be more relevant clusters than the number of recommendations we are testing the metric on. So we make the scores meaningful by taking a minimum of both of these numbers, as the maximum number of relevant clusters you can recommend from is bounded by number of recommendations, if $|R| < |C|$, and in other cases we are bounded by the number of relevant clusters.

The higher the number of unique and relevant clusters represented, the better (more diverse) and this will have a 
 serendipity score of closer to 1. The fewer unique and relevant clusters represented, the closer to 0 (and less diverse) the recommendation list's score will be.

This approach is a simple way to track the extent to which items being recommended to a user are ``spread out"  across their interests. Instead of having the entire recommendation list taken up by a single cluster of items (which could be a sign of echo chambers), recommendation lists with that have serendipity are those that have a wide range of relevant items to the user, even if it is only slightly relevant

\section{Lexicase Selection for Recommendation Systems}
This section outlines proposed methods to augment recommender systems to promote the kinds of recommendations that would result in users breaking out of echo chambers or seeing serendipitous content.

The preservation of serendipity requires more than simply ``novel" items. Truly diverse items do not fit the pattern of what the user has seen in the past. This does not necessarily imply that the best way to select items to tackle echo chambers is to show users items that they are predicted to hate. Indeed, these recommendations do not fit the pattern of what the user has seen in the past, yet they seem to be missing something crucial: a reason to care. To provide this reason, we must find items that the user has a partial interest in, yet are different to what they normally watch.  Oftentimes, these items are lost by the recommender system's preliminary filtering due to their dissimilarity to most items that the user has interacted with. To ensure that these items are not lost from the list, a mechanism to preserve the items that have effective diversity needs to be employed.

In order to preserve this kind of diversity, we can take a page out of the evolutionary computation diversity preservation literature and employ a selection strategy known as lexicase selection. As lexicase selection requires multiple ``objectives" to evaluate on, We use the matching of a user embedding to a \emph{specific dimension} of item space as the evaluation metrics. For matrix factorization, where preference is computed as a dot product between a user and item embedding, we instead find the element wise product between these as a metric of how much this item matches a user's interests on a specific dimension.

Traditionally, recommender systems aggregate user preferences into a single scalar value, which limits the ability to recommend items that match the user's unique preferences. To overcome this, we can use lexicase selection to de-aggregate user preferences into multiple dimensions or "liking" factors. This can be done using any de-aggregation method we choose. In this work, we use a neural network to de-aggregate user preferences into preference features, which are then used as multiple objectives in a multi-objective optimization framework to generate serendipitous recommendations. The sum of these preference features equals the original preference scalar, but the de-aggregation allows for a more nuanced understanding of user preferences and the ability to recommend items that match those preferences in unique ways.

In order to select a single item from these three, we can appeal to the multi-objective selection literature from evolutionary computation. Each of the three preference features can be thought of as an objective that we are trying to maximize with our recommendation list. For this reason, it is a faithful representation of the parent selection problem in evolutionary computation: given a set of individuals that have different scores on different objectives, how do you select $k$ of them to be parents for the next generation. A relatively new approach to solving this problem is known as Lexicase Selection \citep{spector_assessment_2012}, and has been shown to be especially effective for diversity preservation \citep{helmuth_effects_2016}.

In most applications of lexicase selection, the set of items is filtered down by only keeping the items that have \emph{exactly} the best score on each dimension (match on a feature). Due to the fact that our features are often real-valued, it is not appropriate to take the items at each step that have the best possible matching on that feature, as there will likely never be any ties (leading to only one feature ever being used for selection). As we often want interestingly unique combinations of features, we must relax the elite-ness condition that is used in each filtering step. To solve this issue, we adopt a version of lexicase selection that works for real-valued fitness functions known as epsilon lexicase selection \cite{lacava2019}.

The algorithm for lexicase selection for recommendation systems can be found in Algorithm~\ref{alg:lex}\footnote{We implemented a variant of this algorithm for computational efficiency, where we only run the outer loop a set number of times rather than running it for all features in the shuffled list. If you choose the set number high enough you enjoy a gain in computation with a loss of the benefits that come with full lexicase. Through a set of preliminary experiments, we found 10 to be a good bound when the number of features was within 100.} (adapted from \cite{ding2022lexicase}). First, we begin by determining the so-called preference features (the deaggregation of the single preference value that a user has for an item) for every item for a single user. To build a recommendation list for this user, one begins by shuffling the indices of the features into a random order. This ordering will be used to select a single item to be added to a recommendation list. For every feature index in the list, in the order that they were shuffled into, we begin to filter down the set of candidate items by only keeping the items that have within $\epsilon$ of the best match for the user \emph{for that feature}. Note that the item's match on all other features is disregarded at this step. This algorithm has the run time of $\mathcal{O}(N\cdot C)$ where $N$ is the number of features, and $C$ is the number of items to be selected from. This runtime is simply to select a single item. Whilst this seems to be a large number, \citet{helmuth_population_2022} show that in practice, the time complexity is much lower. Despite this, the runtime of lexicase selection still seems to be one of its major disadvantages, although there is active work in this area \citep{ding2022lexicase, hernandez_random_2019}.

\begin{algorithm}[t]
\hrulefill

\vspace{0.5em}
    \KwData{
    \begin{itemize}
        \item \texttt{features} - a sequence of all features (dimensions) of each item to be used for selection (higher is better for a user)
        \item \texttt{items} - the (possibly shortlisted) list of items to be selected from
    \end{itemize}
    }
    \KwResult{
    \begin{itemize}
        \item an item to add to the recommendation list
    \end{itemize}
    }
    \For{\texttt{feature} in \texttt{shuffled\_features}}{
        $i \gets$ the index of \texttt{feature} in \texttt{features} 
        
        \texttt{items} $\gets$ the subset of the current \texttt{items} that has within $\epsilon$ of the best match for this feature.

        \If{\texttt{items} contains only one single item
        }{\KwRet{\texttt{item}}}
    }
    \texttt{item} $\gets$ a randomly selected item in \texttt{items}

    \KwRet{\texttt{item}}
\vspace{0.5em}

\hrulefill
\vspace{0.1em}
    \caption{Lexicase Selection for Item Selection}
    \label{alg:lex}
\end{algorithm}


Using this intuition, we can craft a diverse recommendation for $u_1$ out of the three films present. Say that the first feature in an arbitrary shuffle is feature 3, or romance. Item 2 has the highest score at that dimension (as $0.8*0.8 > 0.8*0.1$ and $0.8*0.8 > 0.8*0.7$), so it will be selected to be shown to the user. Notice that item 2 has a very low match on feature 1, or comedy. $I_2$ will be recommended to $u_1$ despite the fact that the user really likes comedy movies. If we could now perform this selection scheme for all (or perhaps a shortlist) of movies in the entire dataset, it could lead to $u_1$ being shown many movies that match their preference along different combinations of features, which could lead to a very fulfilling and diverse recommendation list.

\section{Benchmarking Recommender Systems} Our experiments were performed using Neural Matrix factorization (NeuMF) \citep{he_neural_2017} as the baseline model. For this system, we compared performing lexicase selection on a de-aggregated preference score for a user-item pair as outlined in the previous section, to simply using the aggregated preference score to select these items. We also compare a mixing method, that mixes both lexicase selection and the baseline together in a 1:1 ratio. As a baseline, we include a random selection, where instead of using lexicase selection to select from our list of movies, we simply randomly select items.

Hyperparameter tuning was done using a random search on all hyperparameters, training the model on 70\% of the data and testing it on the rest using Mean Absolute Error as the performance evaluation of the model. We use the MovieLens-1M\footnote{\url{https://grouplens.org/datasets/movielens/1m/}} dataset to build our recommendations.

We compare the performance of different final list creation methods for NeuMF tasked with making top-5, 10, 15, 20, and 25 recommendations. Here top-$k$ recommendation is just selecting $k$ best items for the user using the appropriate method as described earlier. We use four different metrics to evaluate the performance of each of our recommendation systems: Coverage, Personalization, Hit Rate, and our above proposed set difference metric. We compare the recommendation system with rankings (r), lexicase (l), a 50\% mix of both (m-50) or a random selection (random) baseline. The mix was conducted by a simple interleaving of a list generated with rankings, and a list generated with lexicase.  The code used for these experiments is hosted online \footnote{
\href{https://colab.research.google.com/drive/1iyk1Ia_ptuyzybRrAjAZmucnh_WDZMoH?usp=sharing}{Google Colab Link for Experiment code}
}.


\section{Results and Discussion}

\begin{figure}
     \centering
     \begin{subfigure}[b]{0.45\textwidth}
         \centering
        \includegraphics[width=1\textwidth]{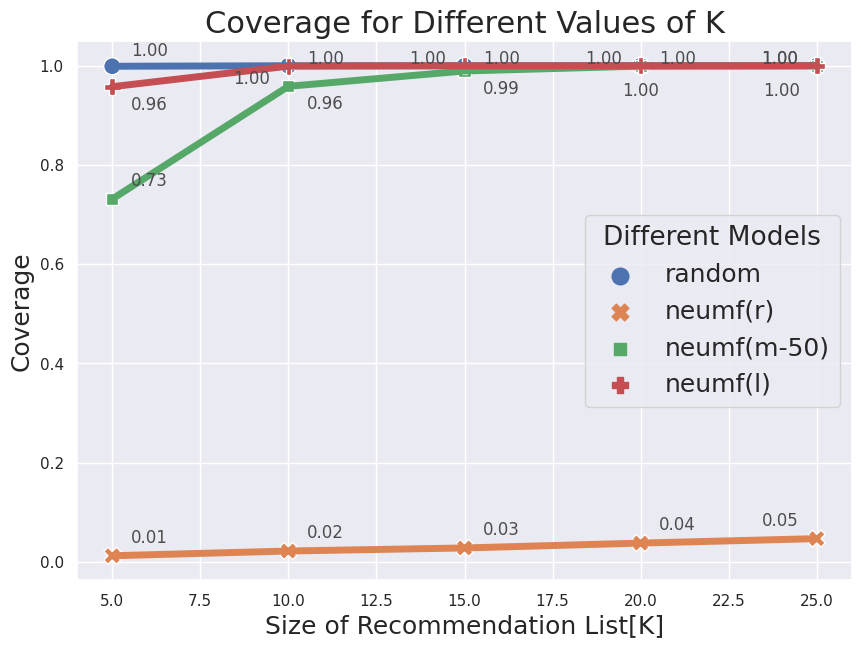}
        \caption{Coverage}
        \label{fig:Coverage}
     \end{subfigure}
     \hfill
     \begin{subfigure}[b]{0.45\textwidth}
        \centering
        \includegraphics[width=1\textwidth]{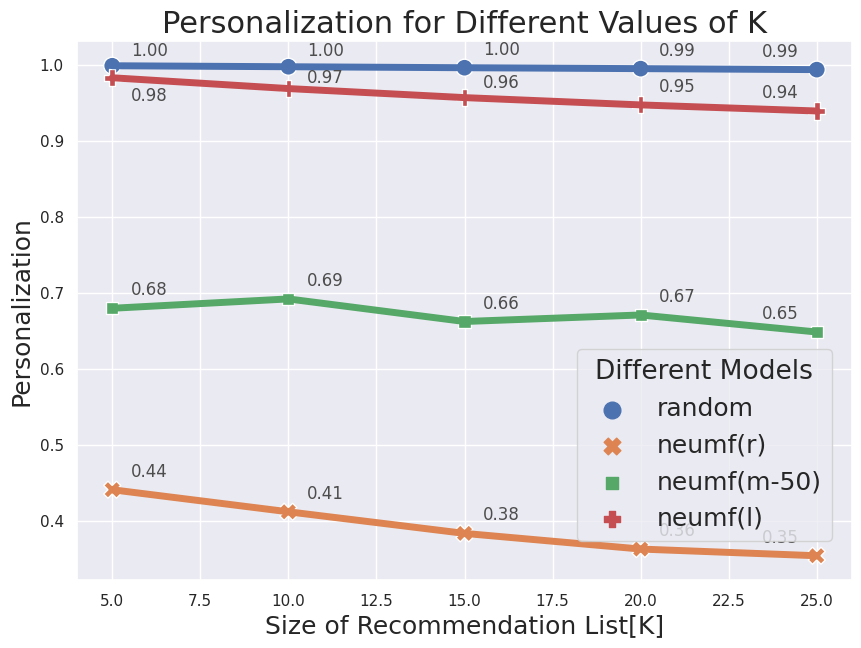}
        \caption{Personalisation}
        \label{fig:Personalisation}
     \end{subfigure}\\
     \begin{subfigure}[b]{0.45\textwidth}
         \centering
        \includegraphics[width=1\textwidth]{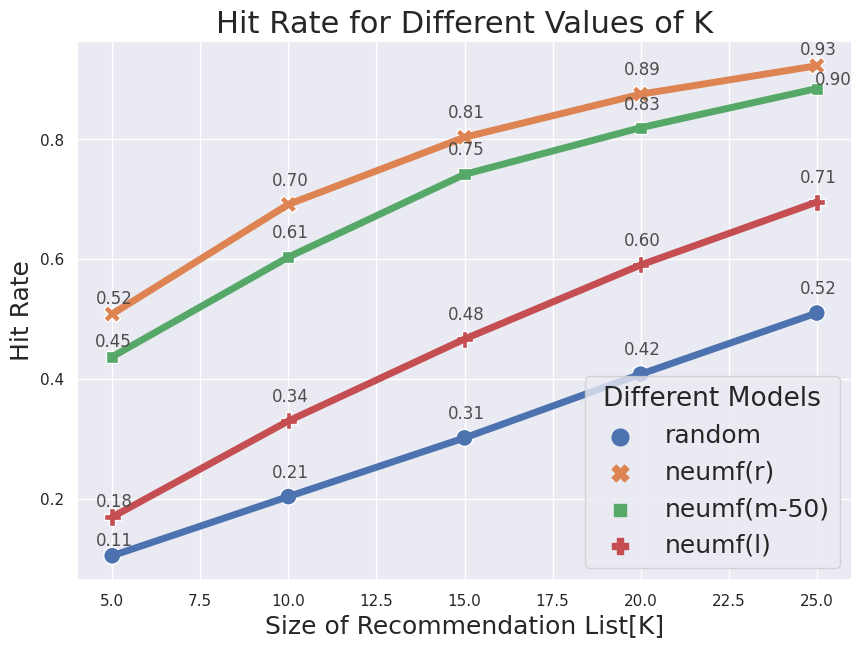}
        \caption{Hit Rate}
    \label{fig:HitRate}
     \end{subfigure}
     \hfill
     \begin{subfigure}[b]{0.45\textwidth}
         \centering
        \includegraphics[width=1\textwidth]{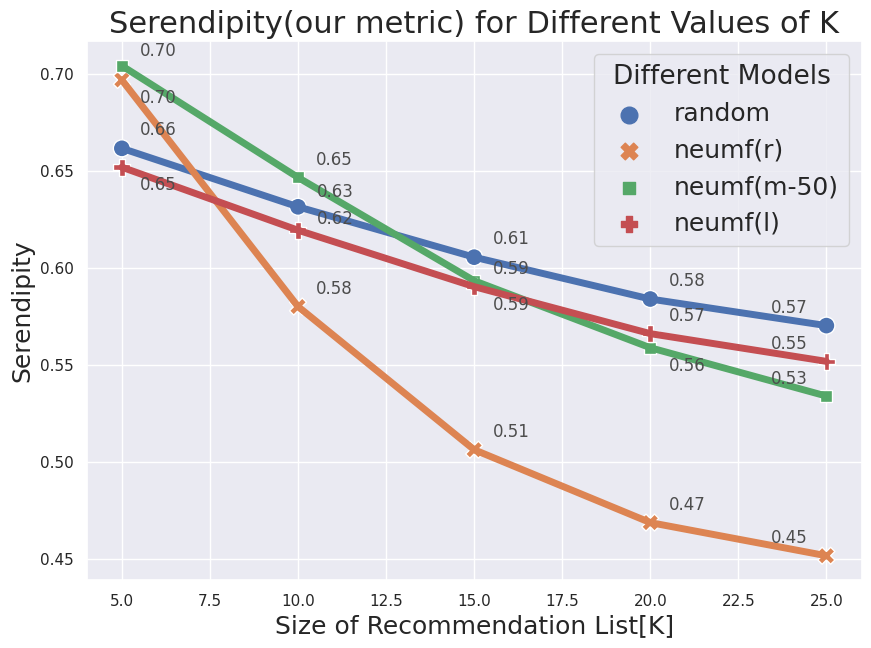}
        \caption{Set difference serendipity (our proposed metric)}
        \label{fig:Diversity}
     \end{subfigure}
        \caption{A variety of metrics comparing NeuMF performed with the aggregated preference score (r), lexicase selection on the de-aggregated preference scores (l) or a mix of both (m-50). We look at the Top-K recommendations for a variety of values of $k$ for each of these metrics.}
        \label{fig:three graphs}
\end{figure}

The results for Coverage, Personalization, Hit Rate and our proposed Serendipity metric can be found in Figure~\ref{fig:three graphs}. We can see that lexicase selection or a mixed method outperforms the ranking based-approach using all four of the metrics. We will now discuss each metric in turn.

The coverage metric (Figure ~\ref{fig:Coverage}) measures the proportion of the items that are recommended for at least a single user. When it comes to coverage, the higher the better. We see that the variants using lexicase selection perform better than the respective versions using a ranking-based approach. This could be likely due to lexicase selection helping select items that have high matching in some dimensions, but are low matching in others, and such would not be selected if you considered an aggregate measure. When using a de-aggregated measure, they are selected more often, which could explain the increase in coverage. With respect to the random selection baseline, we see that lexicase selection and mixed both rapidly approach the same performance as random. A reason for random selection resulting in high coverage is due to the relatively few number of users and movies in this dataset. Even with just 5 movies in each recommendation list, it is likely that all movies appear in atleast one recommendation list

The second metric analyzed was personalisation (Figure~\ref{fig:Personalisation}). This is the metric that measures how different the recommendations are for each user. As with coverage, the higher the personalization, the better. This is because we want each user's recommendations to be less related to one another, implying the lack of echo chambers. We can see here again that Lexicase selection and the mixed approach also outperform the ranking based approach. This could be due to lexicase selection's ability to specifically target features that have a high degree of matching for a specific user, without regarding the features that the user might not like. This could lead to more personalized recommendations as users are shown things that are more specifically tailored to their interests in interesting and unique ways.

The third metric analyzed was hit rate (Figure ~\ref{fig:HitRate}). This metric measures the ability of a recommender system to accurately predict which items a user also has rated in the dataset. This metric can be used as a form of ``accuracy" metric. For hit rate, as one would expect, we see that the ranking based approach outperforms both the lexicase and mixed strategies. This is the expected result as we are utilizing lexicase selection to help preserve the diversity of the recommendations whilst not really attempting to improve the hit rate. 

It is also important to note that metrics like hit rate are used to measure how well the recommendation system does at predicting items that users have interacted with \emph{in the data set}. This data set might have been generated while using a separate recommendation system, and so the results using hit rate and similar metrics are biased towards recommendation systems that are similar to those used to generate the original data set. With our lexicase selection based systems, the recommendations perhaps do not align with the data set, and so is not a very faithful comparison.

The final metric used is our own serendipity metric using set differences (Figure~\ref{fig:Diversity}). We can see here that for small $k$, mixing lexicase selection with the ranking based approach performs the best. As $k$ increases, however, the serendipity score decreases to a point where regular lexicase outperforms the other two. Random selection performs reasonably well despite its simplicity on this benchmark. This can be explained by the low hit rate that random achieves. Because random does not capture many relevant clusters, the ratio of unique relevant clusters to relevant clusters is high. A possible reason for our serendipity score decreasing with increase $k$ is that it measures the maximum attainable diversity given a specific recommendation list $R$ and value of $k$. As the size of $k$ increases, it is harder to fill up the list with unique movies, and so the proportion of movies that are unique decrease. This is intuitive as the larger the list of recommended items, the higher the chance that you'll see multiple movies that are very similar to each other. As we can see that the method that mixes lexicase selection and ranking outperforms all other approaches for small $k$, we are effectively maintaining serendipity when there are not many movies in the list. However, as that list increases in size, the ranking based approach fills up more and more of the list with similar items, and so collapses the serendipity score for high $k$. Lexicase selection and the mixed version have a much less steep drop off of this value due to its diversity maintenance capacity.




\section{Conclusion}
We proposed a serendipity metric to specifically target echo chambers and homophily in recommendation systems. This metric uses a set difference approach in order to determine how diverse the recommendations are, when taking into account a user's many different interests.

We then proposed an augmentation to a recommendation system based on diversity preservation work in evolutionary computation. We propose to use a de-aggregated form of preference prediction, which is predicting preference for certain item \emph{features} instead of items themselves. Then, we use a form of parent selection known as Lexicase selection to pick the items that have a diverse yet high matching set of features to those that a user is interested in.

We compared two different recommendation systems, Neural Collaborative Filtering and Neural Matrix Factorization with regular aggregated ranking, de-aggregated lexicase selection, and also a mix of the two. We found that lexicase selection outperforms the ranked counterpart across both systems, a variety of recommendation list sizes, and also four different metrics. The mixed method outperforms lexicase selection on hit rate, and on our serendipity metrics for some values of $k$.

This work highlights that lexicase selection or other diversity preservation techniques from the evolutionary computation literature are able to be applied to recommendation systems in order to help improve the diversity, move people away from echo chambers, and also help decrease the amount of homophily in the recommendations. The conclusions from this work open many different pathways for future research. A possible area of study is to directly address homophily by using lexicase selection on the user demographic context, as opposed to just their embeddings. This will help lexicase selection select items that are diverse for a person \emph{given} who they are and their background. Future investigation would also benefit from using lexicase selection to its full capacity, as opposed to limiting to just 10 features at most being used for selection. Finally, the application of other diversity preservation techniques from the evolutionary computation literature, such as Quality Diversity, would be a very interesting area to move into.
\newpage
 \bibliography{references}  
\newpage

\end{document}